\documentclass[aps,prl,twocolumn,10pt,superscriptaddress]{revtex4-1}
\usepackage{amssymb,amsthm,amsmath,amsfonts}
\usepackage{xcolor,graphicx,ulem}
\usepackage[colorlinks=true,urlcolor=blue,citecolor=blue,linkcolor=blue]{hyperref}

\begin{document}

\title{Incomplete Detection of Nonclassical Phase-Space Distributions}

\author{M. Bohmann}\email{martin.bohmann@uni-rostock.de}
	\affiliation{Arbeitsgruppe Theoretische Quantenoptik, Institut f\"ur Physik, Universit\"at Rostock, D-18051 Rostock, Germany}
\author{J. Tiedau}
	\affiliation{Integrated Quantum Optics Group, Applied Physics, University of Paderborn, 33098 Paderborn, Germany}
\author{T. Bartley}
	\affiliation{Integrated Quantum Optics Group, Applied Physics, University of Paderborn, 33098 Paderborn, Germany}
\author{J. Sperling}
	\affiliation{Clarendon Laboratory, University of Oxford, Parks Road, Oxford OX1 3PU, United Kingdom}
\author{C. Silberhorn}
	\affiliation{Integrated Quantum Optics Group, Applied Physics, University of Paderborn, 33098 Paderborn, Germany}
\author{W. Vogel}
	\affiliation{Arbeitsgruppe Theoretische Quantenoptik, Institut f\"ur Physik, Universit\"at Rostock, D-18051 Rostock, Germany}

\date{\today}

\begin{abstract}
	We implement the direct sampling of negative phase-space functions via unbalanced homodyne measurement using click-counting detectors. 
	The negativities significantly certify nonclassical light in the high-loss regime using a small number of detectors which cannot resolve individual photons.
	We apply our method to heralded single-photon states and experimentally demonstrate the most significant certification of nonclassicality for only two detection bins.
	By contrast, the frequently applied Wigner function fails to directly indicate such quantum characteristics for the quantum efficiencies present in our setup without applying additional reconstruction algorithms.
	Therefore, we realize a robust and reliable approach to characterize nonclassical light in phase space under realistic conditions.
\end{abstract}
\maketitle

\paragraph*{Introduction.---}\hspace{-3ex}
	Photons embody the wave-particle dualism---a governing principle of quantum physics---as they represent the elementary particle of electromagnetic waves.
	They also provide the essential building blocks for scalable quantum technologies \cite{Northup14,Koenderink15}, and fundamental quantum effects are intertwined with the properties of such particles of light, e.g., antibunching \cite{Kimble77} and sub-Poisson photon statistics \cite{Short83}.
	For this reason, resource-efficient detection schemes which enable the characterization on the single-photon level are indispensable.

	In order to uncover quantum features of a physical system, the concept of phase-space distribution functions has been extended from the classical domain to be applicable in quantum optics.
	This yields the prominent Wigner function \cite{Wigner32} and its $s$-parametrized generalizations \cite{Cahill69}, which include the Glauber-Sudarshan $P$ \cite{Glauber63,Sudarshan63} and Husimi $Q$ \cite{Husimi40} functions.
	Based on such quasiprobability distributions, quantum effects can be uniquely identified via negativities within them \cite{Titulaer65,Hudson74,Mandel86}.
	Thus, since its first applications \cite{Smithey93,Leibfried96}, the reconstruction of the Wigner function became a frequently applied method.
	This led to remarkable insights into the quantum properties of complex quantum systems, such as reported in Refs. \cite{Deleglise08,Vlastakis13,Morin14,Wang16,Sychev17}.
	Still, there are limitations to this approach.
	For example, the directly reconstructed Wigner function of a single photon becomes entirely non-negative when the overall loss exceeds $50\%$ \cite{Lvovsky01}---a challenging bound for many experimental scenarios.
	In such cases, negativities can be recovered by employing optimized reconstruction algorithms which compensate for losses \cite{Lvovsky04}.

	In addition, information-complete measurements and data processing algorithms are necessary for a phase-space quantum state reconstruction.
	For instance, the reconstruction can be based on quadrature \cite{VogelBook,Lvovsky09} or displaced photon-number measurements \cite{Wallentowitz96}.
	The latter can be obtained via unbalanced homodyne detection \cite{Wallentowitz96}, which has been implemented, e.g., in Refs. \cite{Banaszek99,Laiho10,Donati14}.
	However, in many practical scenarios, such optimal detection schemes are not accessible because of experimental limitations, such as nonunity detection efficiencies and a limited resolution of adjacent photon numbers.
	In such cases, an inversion from the measured data to the photon-number statistics is required \cite{Banaszek99,Laiho10,Donati14}, which is problematic when employing information-incomplete detection schemes.
	It is also noteworthy that even a partial state reconstruction can reveal nonclassicality as it was demonstrated for marginal distributions \cite{Park17}.

	A more realistic resource is quasi-photon-number-resolving or click-counting detectors \cite{Rehavcek03,Achilles03,Fitch03,Schettini07,Lundeen09,Matthews16}, which discriminate between clicks rather than photocounts.
	Despite their limitations, a number of vital quantum features can be directly obtained from the recorded click-counting statistics \cite{Bartley13,Sperling15,Sperling16,Bohmann17}.
	Furthermore, the theory of phase-sensitive measurements with such detectors has been derived \cite{Sperling15a,Luis15,Lipfert15}.
	Specifically, the click-based counterpart to the unbalanced homodyne detection has been proposed together with a click-based versions of phase-space functions \cite{Luis15}.
	Yet, an experimental confirmation of this theory is missing to date.

	In this Letter, we report on the direct sampling of phase-space functions of heralded single-photon states via unbalanced homodyne detection with time-bin-multiplexed click-counting detectors.
	The resulting phase-space functions directly show negativities even for quantum efficiencies significantly below $50\%$.
	Moreover, this is achieved via a direct sampling of the measured click statistics, which does not require an inversion to the photon-number statistics and is intrinsically robust to losses.
	We also confirm the theoretical prediction from Ref. \cite{Luis15} that fewer detection bins lead to more significant signatures of nonclassical light.
	Therefore, we implement a reliable method to access the quantum properties of light for application under realistic conditions.

\begin{figure}[tb]
	\centering
	\includegraphics[width=\columnwidth]{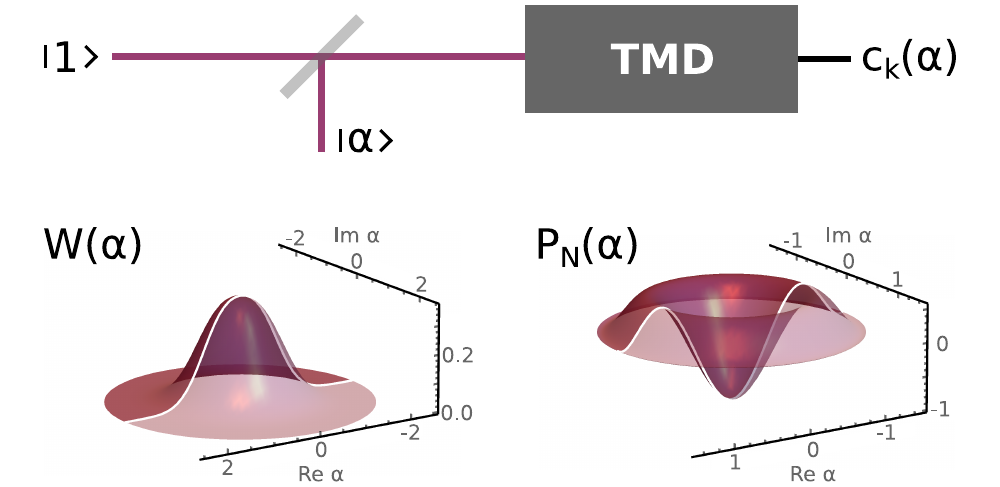}
	\caption{
		Top:
		A signal, e.g., a single photon $|1\rangle$, and a coherent state $|\alpha\rangle$ are superimposed on a highly transmissive beam splitter and measured with a click-counting device, which is in our case a time-bin multiplexing detector (TMD).
		Bottom:
		The non-negative Wigner function $W(\alpha)$ for single photon and a detection efficiency of $21\%$ (left) is depicted.
		In the same scenario, its directly sampled click-counting counterpart (right, $N=4$) shows clear negativities.
	}\label{fig:1}
\end{figure}

\paragraph*{Click-based phase-space functions.---}\hspace{-3ex}
	In unbalanced homodyne detection \cite{Wallentowitz96}, the quantum light under study is displaced by mixing it on an unbalanced beam splitter with a weak coherent state.
	The resulting displaced state is then measured with a photon-number-resolving detector.
	Recently, the theory for such a scheme was generalized to allow for an operation using click-counting devices \cite{Luis15}; see Fig. \ref{fig:1}.

	In such a scheme, the directly measured click statistics $c_k(\alpha)$ is the probability that $k$ out of the $N$ detection bins record coincidental clicks \cite{Sperling12}.
	Based on this statistics, we obtain---up to a positive normalization constant---the desired phase-space function \cite{Luis15},
	\begin{equation}\label{eq:PN}
		P_{N}\left(\alpha;x\right)=\sum_{k=0}^N\left(\frac{x-2}{x}\right)^k c_k\left(\alpha\right),
	\end{equation}
	for any even number $N$ of detection bins (the parameter $x$ is discussed in the next paragraph).
	In addition, the sampling error is given by
	\begin{align}\label{eq:PNerror}
		\left[\Delta P_N \left( \alpha ; x \right)\right]^2 = \sum_{k=0}^N 
		\left(\frac{x-2}{x} \right)^{2k} \frac{c_k (\alpha ) \left[ 1- c_k (\alpha ) \right] }{\omega},
	\end{align}
	where $\omega$ is the number of recorded data points for a given displacement $\alpha$.

	The parameter $x$ in Eq. \eqref{eq:PN} can be related to $s$-parametrized phase-space functions \cite{Cahill69} ($s\in[-1,1]$) and the detection efficiency $\eta$ via $x=\eta (1-s)$.
	Let us recall that $s=0$ defines the Wigner function.
	As demonstrated in Ref. \cite{Luis15}, any negativity in $P_{N}\left(\alpha;x\right)$ for any even $N$ is a sufficient condition for nonclassicality.
	In the limit of an infinite number of detection bins, $N\to\infty$, $P_{N}\left(\alpha;x\right)$ approaches the true $s$-parametrized phase-space function.
	This then results in a necessary and sufficient nonclassicality characterization.
	Also note that beyond the specific identification $x=\eta(1-s)$, an arbitrary nonzero number can be assigned to the parameter $x$ \cite{SM}.

	One of the main benefits of the phase-space distribution Eq. \eqref{eq:PN} for a finite and even $N$ is that it can become negative even if its $s$-parametrized counterpart (without corrections for detection losses) is completely non-negative \cite{Luis15}.
	To illustrate this, the example of a single photon is shown in Fig. \ref{fig:1} for a quantum efficiency $\eta=21\%$, which corresponds to our experimental conditions.
	Clearly, the phase-insensitive Wigner function ($s=0$) is non-negative.
	Still, the theory of the click-counting phase-space function for $x|_{s=0}=\eta$ predicts clear negativities, $P_N(0;x)<0$.

\begin{figure*}[tb]
	\centering
	\includegraphics[width=1.5\columnwidth]{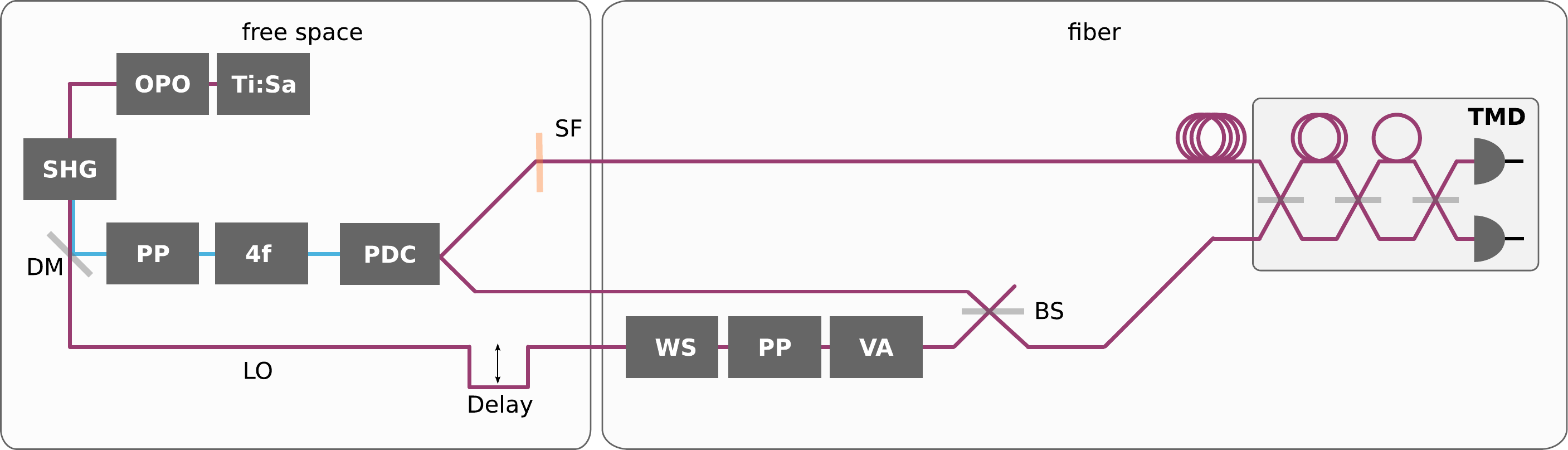}
	\caption{
		Schematic experimental setup.
		A Ti:sapphire laser (Ti:Sa) (down-sampled to $1\,{\rm MHz}$) drives an optical parametric oscillator (OPO).
		The pulsed light is frequency doubled by second harmonic generation (SHG), and the pump is filtered with a dichroic mirror (DM).
		The SHG pulse undergoes pulse picking (PP) down to $1\,{\rm MHz}$, followed by a 4-f (4f) line to set the pulse length, and is used to generate photon pairs in a parametric down-conversion (PDC) process.
		Before both modes are recorded with time-multiplexed detectors (TMD), one of them is coherently displaced by the local oscillator (LO), mode matched by a time delay, spectral wave shaper (WS) and pulse picker (PP), and attenuated using a variable attenuator (VA), before interfering at an asymmetric beam splitter (BS), with reflection-transmission ratio of $17:83$.
	}\label{fig:exp}
\end{figure*}

\paragraph*{Experimental implementation.---}\hspace{-3ex}
	The full experiment was conducted using the setup shown in Fig.~\ref{fig:exp}.
	This consists of three main parts: generation of heralded single-photon states, interference with a weak coherent beam, and detection on a time-multiplexed detector (TMD).
	The initial light from a Ti:sapphire oscillator at $80\,{\rm MHz}$ pumps an optical parametric oscillator (OPO) to generate light at $1550\,{\rm nm}$.
	This pumps a periodically poled lithium niobate (PPLN) crystal to generate second harmonic light (SHG) at $775\,{\rm nm}$.
	This light is split from the residual pump using a dichroic mirror, and then used to pump the source of heralded single photons, namely parametric down-conversion in a periodially poled KTP waveguide.
	The source is described in detail in Refs. \cite{Eckstein2012,Harder2013}.
	Prior to the waveguide, the pump is frequency-picked down to $1\,{\rm MHz}$.
	This is required to ensure that there are no overlapping pulses registered by the TMD.
	The temporal mode of this light is then adjusted using a 4-f line, to ensure the bandwidth of the pump is set to give a spectrally decorrelated down-conversed state.
	Meanwhile, the residual pump from the SHG also undergoes pulse-picking and spectral mode shaping such that it is synchronized and mode matched to the down-conversion.
	The overlap is $70\%$, determined from a Hong-Ou-Mandel-type experiment between the heralded single photon and the weak local oscillator.
	The coherent state amplitude $|\alpha|$ is adjusted with a computer-controlled variable attenuator.
	The experimental uncertainty in determining $|\alpha|$ ($\sim20\%$) predominantly arises for temporal drifts in the setup over an experimental run.
	Following interference, the resulting states are incident on a TMD with a bin separation of $100\,{\rm ns}$.
	Heralding is performed by delaying the herald mode by $400\,{\rm ns}$, which is then incident on the other input port of the TMD.

	The advantage of time-bin multiplexing is that we only need two detectors, which in our case are superconducting nanowire single-photon detectors, characterized in a previous experiment \cite{Bohmann17a}, together with a given number of fiber-loop delays to obtain $N/2$ time bins per detector, which correspond to $N$ physical detectors in the spatial multiplexing configuration \cite{Achilles03,Rehavcek03,Fitch03}.
	Our TMD divides an incident light pulse into up to $N=8$ time bins per input.

	Following interference, one mode is measured by a TMD to provide the click statistics $c_k\left(\alpha\right)$.
	To sample the phase-space function $P_N$, we record approximately $\omega=1.4\times10^6$ heralded photon states per coherent amplitude for $N=8$ detection bins.
	In order to achieve fewer bins $(N=2,4)$, we can join several detection bins into clusters \cite{SM}.
	A detailed theoretical model of the state preparation and the measurement is also provided in the Supplemental Material \cite{SM}.

\paragraph*{Direct sampling of phase-space functions.---}\hspace{-3ex}
	Applying Eqs. \eqref{eq:PN} and \eqref{eq:PNerror}, we can directly sample the phase-space function from our data.
	Among other imperfections, our model \cite{SM} takes unavoidable higher-photon number contributions of the heralded single-photon state into account.
	Also, impurities originating from the mode mismatching of the signal and LO are considered, which yields a displacement-dependent dark-count rate \cite{Lipfert15,SM}.
	Finally, let us stress that our estimated overall detection efficiency, obtained via the fit to our model, is $\eta=21\%$---thus, clearly below $50\%$.

\begin{figure}[tb]
	\centering
	\includegraphics[width=\columnwidth]{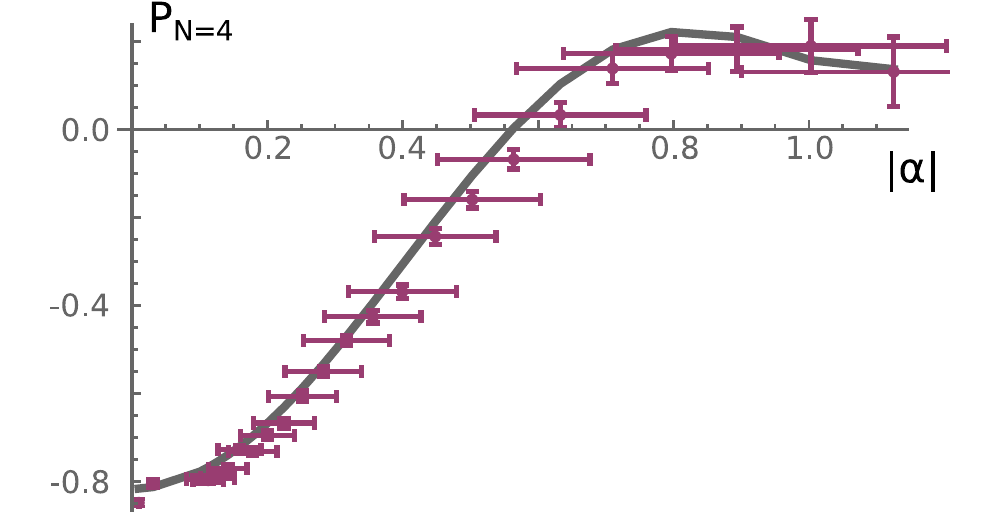}
	\caption{
		The pointwise sampled phase-space function $P_{N}\left(\alpha;x\right)$, for $N=4$ and $x=0.21$, is shown and compared with the theoretical model (solid curve).
	}\label{fig:P41}
\end{figure}

	For the four-bin detection $(N=4)$, our rotationally symmetric, sampled phase-space distribution $P_N$ is shown for $x=\eta$ in Fig. \ref{fig:P41}.
	The most important observation is that this function shows clear negativities around the origin of phase space, certifying the nonclassical character of the quantum state.
	Such a negative dip at $|\alpha|=0$ is typical for single-photon states, cf. Fig. \ref{fig:1}.
	Furthermore, we find a good agreement of the obtained phase-space function with our theoretical model.
	Note that the estimation of the amplitude $|\alpha|$ of the displacement is challenging because the mean photon number cannot be simply obtained by click-counting devices \cite{Sperling12}.
	More importantly, however, this error does not affect the significance of the certified negativities just their location in phase space.

	As mentioned earlier (see also Fig. \ref{fig:1}), the overall detection efficiency of the setup does not allow for a direct reconstruction of a negative Wigner function.
	This would require a postprocessing to refit the data to a lossless scenario \cite{Lvovsky04}.
	With our technique, such postprocessing becomes superfluous since we directly sample our nonclassical phase-space function according to Eq. \eqref{eq:PN}.
	In particular, it is sufficient to find one $x$ for which negativities become significant.
	Such an $x$ can be chosen arbitrarily and without prior knowledge about losses.
	Therefore, our technique is a more direct and robust approach to uncover nonclassical features of light.
	Moreover, the reconstruction of the Wigner function with unbalanced homodyne detection requires a full photon-number resolution \cite{Wallentowitz96}.
	Again, we demonstrate that such a premise is not required with our alternative approach.

\paragraph*{Number-of-bins dependence.---}\hspace{-3ex}
	A surprising theoretical finding in Ref. \cite{Luis15} is that fewer detection bins lead to an improved verification of nonclassicality.
	Let us challenge this statement by comparing our results for different numbers of detection bins.

\begin{figure}[tb]
	\centering
	\includegraphics[width=\columnwidth]{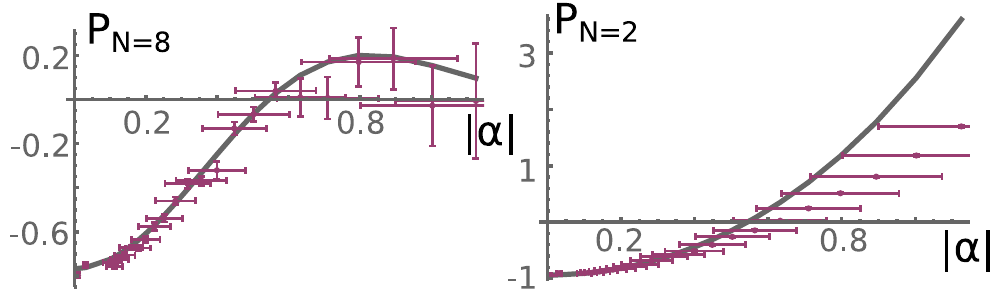}
	\caption{
		The sampled phase-space functions and the theoretical models (solid curves) for eight (left) and two (right) detection bins are displayed.
	}\label{fig:PN42}
\end{figure}

	For this reason, we analyzed the data for two and eight detection bins in addition to the previously studied case, $N=4$.
	The correspondingly sampled phase-space functions are shown in Fig. \ref{fig:PN42} ($x=0.21$) together with the theoretical model.
	Again, theory and experiment exhibit good agreement, and in both cases, $N=8$ and $N=2$, a negativity at the origin of phase space is present.

\begin{figure}[h]
	\centering
	\includegraphics[width=\columnwidth]{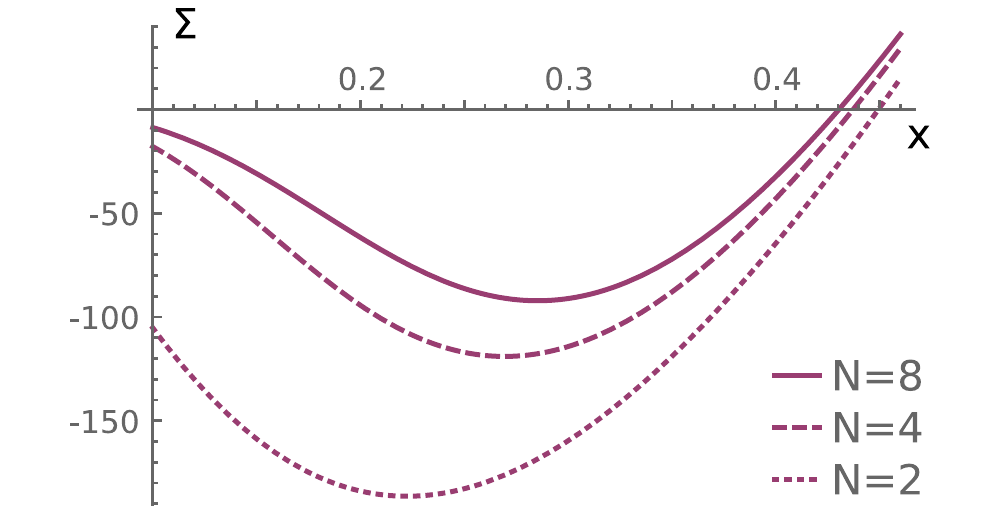}
	\caption{
		The signed significance $\Sigma$ [Eq. \eqref{eq:Sigma}] at the origin ($\alpha=0$) is shown as a function of $x$.
	}
	\label{fig:significances}
\end{figure}

	To quantitatively assess the quality of the verified nonclassicality, we analyze the signed significance $\Sigma$ of the phase-space functions.
	The signed significance is the ratio of the sampled values and their errors,
	\begin{align}\label{eq:Sigma}
		\Sigma=\frac{P_N(\alpha;x)}{\Delta P_N(\alpha;x)},
	\end{align}
	cf. Eqs. \eqref{eq:PN} and \eqref{eq:PNerror}.
	This means that for $\Sigma<0$, we certify the nonclassicality with $|\Sigma|$ standard deviations.
	In Fig. \ref{fig:significances}, the signed significances for the different numbers of detectors are plotted as a function of the parameter $x\in[0.10,0.45]$.
	We find that the curve for $N=4$ (dashed) is always below the one for $N=8$ (solid), and the case $N=2$ (dotted) is below both.
	Therefore, we confirm for our data that the verification of nonclassicality is indeed improved by reducing the number of detection bins.

	Furthermore, the maximal verification of nonclassicality with $|\Sigma|=186$ standard deviations is obtained for two detection bins and $x=0.22$.
	This corresponds approximately to the click-counting phase-space distribution related to the Wigner function, where $x=\eta(1-s)$ and $x|_{s=0}=0.21$.
	Surprisingly, the most significant negativities for the four- and eight-bin scenarios are obtained for higher $x$ values, corresponding to phase-space functions closer to the $Q$ function ($s=-1$).
	However, all distributions eventually become non-negative for larger $x$ values.
	It is also worth mentioning that $x\to0$ corresponds to the often highly singular Glauber-Sudarshan $P$ function $(s=1)$.
	In this limit, we do not obtain significant negativities from the sampled data, $\Sigma\gtrsim 0$.

\paragraph*{Discussion.---}\hspace{-3ex}
	Let us briefly compare our technique to other approaches.
	The advantage of our method is a direct sampling rather than employing challenging optimization algorithms.
	Thus, it allows one to characterize quantum light in a simple and direct way.
	Furthermore, our approach works even in the high-loss regime.
	In particular, it is sufficient to find one $x$, which we can freely choose without imposing or estimating the overall losses in our setup, such that a negative phase-space distribution is obtained, cf. Fig. \ref{fig:significances}.
	Other data processing approaches, e.g., based on balanced homodyne detection, can also correct for losses.
	Yet, they require precise information about the losses for their operation.
	Other optimization strategies in data analysis, such as used maximum likelihood approaches, also become superfluous with our technique.
	Therefore, our sampling method is fast and robust and requires minimal information about the detection scheme.

\paragraph*{Conclusions.---}\hspace{-3ex}
	We implemented a detection scheme which yields nonclassical phase-space distributions via a direct sampling from the data obtained by an information-incomplete detection system.
	Our technique renders it possible to certify nonclassical light under realistic conditions even for high losses.
	Furthermore, sophisticated reconstruction algorithms become superfluous.
	The application to a heralded single-photon state in the presence of high losses also confirms the theoretical prediction that fewer detection bins can be even advantageous for the verification of quantum light.

	Our technique resembles an unbalanced homodyne detection scheme which, however, uses widely accessible click-counting devices instead of photon-number-resolving detectors.
	The measured click-counting distribution is used to sample our desired function by scanning over the phase space, which is achieved by adjusting the local oscillator.
	Our click-counting device is a time-bin multiplexing detector which allows us to resolve up to eight bins measured with two detectors.
	This equals eight physical detectors when compared to other multiplexing approaches.

	Using a parametric down-conversion source, we produced single-photon states, and a theoretical model was developed.
	Our analysis showed that we operate in a comparably high-loss regime, for which other methods would require loss corrections in order to exhibit the nonclassicality.
	Here, however, we have been able to uncover nonclassical light with high statistical significance via direct sampling.
	This shows that imperfect measurement can offer highly sensitive nonclassicality tests, even compared to perfect photon-number-resolving detectors.

	Beyond our proof-of-concept demonstration for phase-symmetric single-photon states, our technique can be applied to arbitrary quantum light and straightforwardly generalized to multimode scenarios.
	Because of its simplicity and reliability, our method has the potential to find many applications in quantum optics and quantum technology, which require sturdy components and directly accessible methods.
	Therefore, we realized a technique which provides a practical tool for the characterization of nonclassical light in phase space.

\paragraph*{Acknowledgements.---}\hspace{-3ex}
	M.B. gratefully acknowledges financial support by the Deutsche Forschungsgemeinschaft through Grant No. VO 501/22-2.
	J.T., J.S., C.S., and W.V. acknowledge funding from the European Union's Horizon 2020 research and innovation programme under Grant No. 665148 (QCUMbER).
	C.S. acknowledges funding from the Gottfried Wilhelm Leibniz-Preis.
	T.B. acknowledges financial support from the DFG (Deutsche Forschungsgemeinschaft) under SFB/TRR 142.



\onecolumngrid
\includegraphics[width=\textwidth]{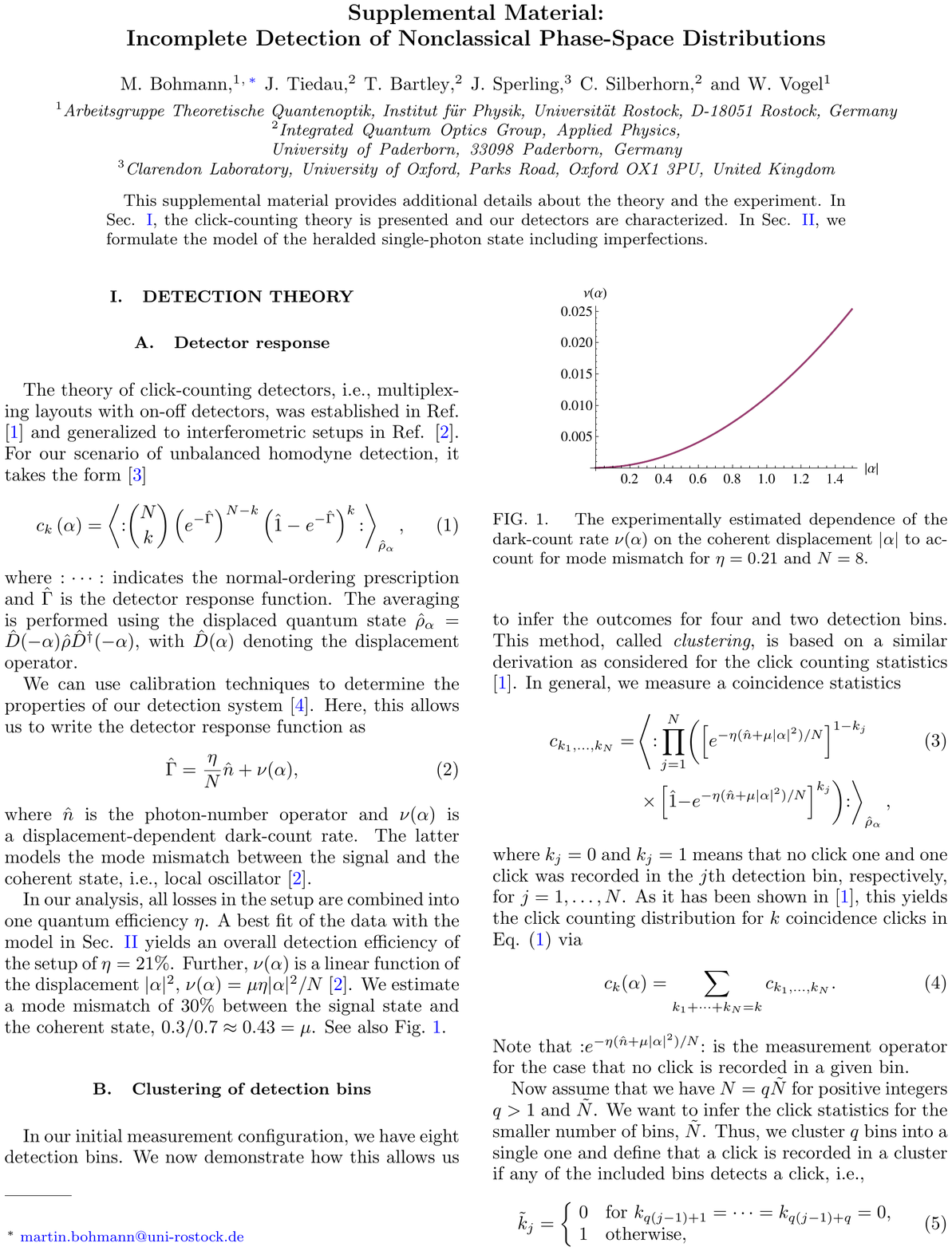}

\includegraphics[width=\textwidth]{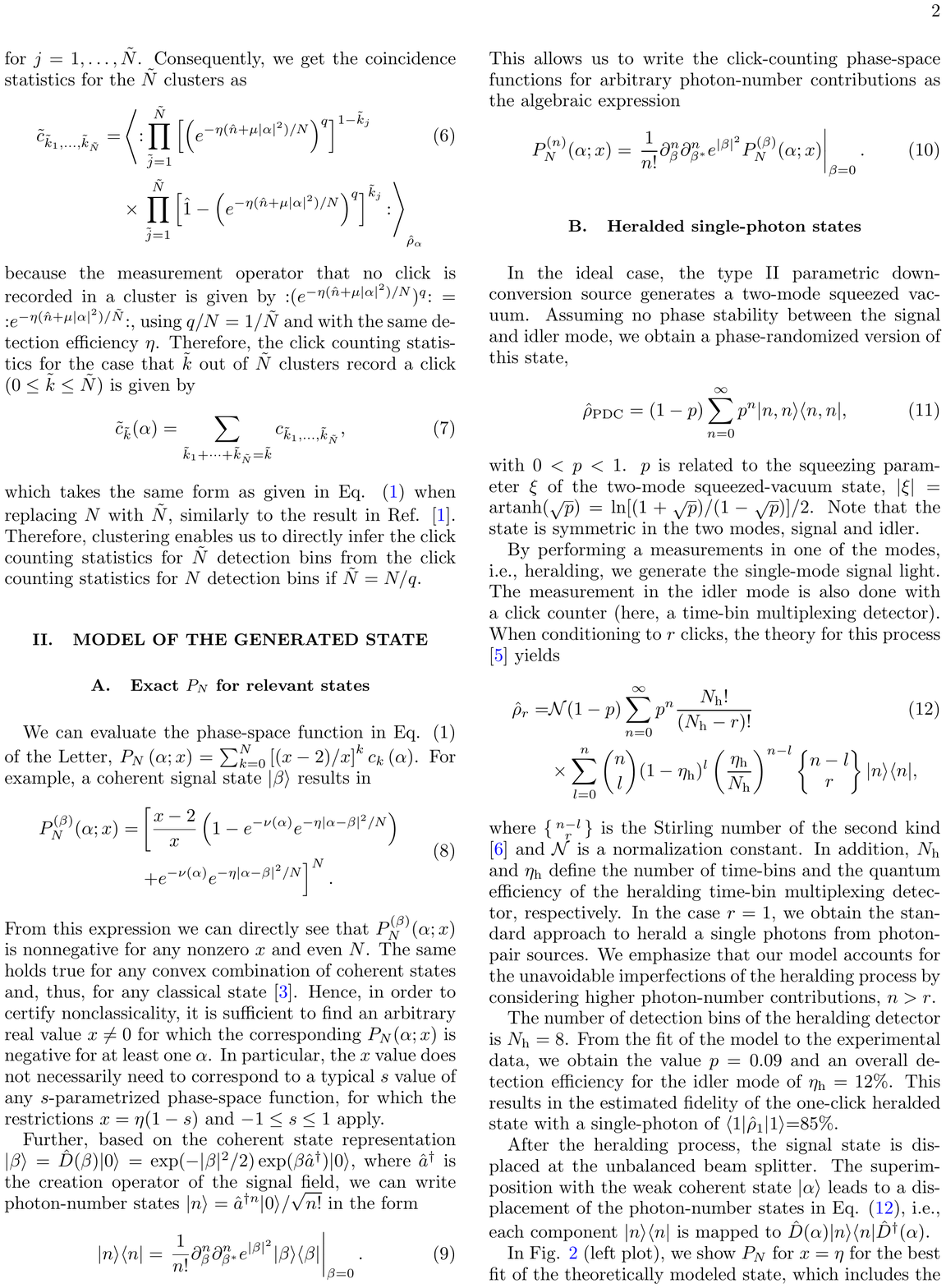}

\includegraphics[width=\textwidth]{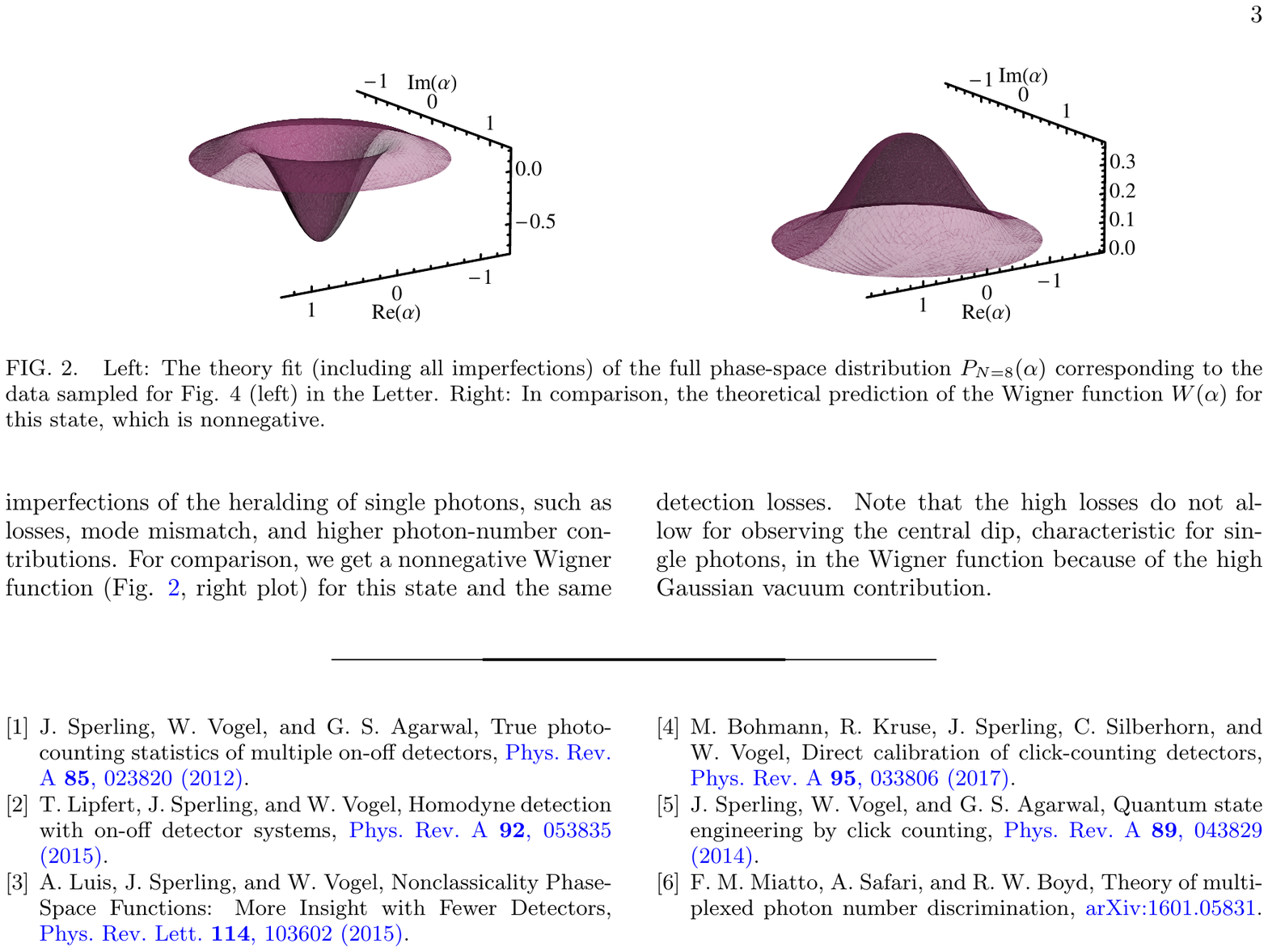}

\end{document}